\documentclass[sigconf]{acmart}
\usepackage[utf8]{inputenc}
\usepackage{booktabs}
\usepackage{listings}
\usepackage{graphicx}
\usepackage{subcaption}
\usepackage{import}
\usepackage{algorithm}
\usepackage{algorithmic}

\renewcommand{\algorithmiccomment}[1]{\bgroup\hfill$\rhd$~#1\egroup}
\newcommand{\helium}[0]{\textit{FuzzerAid}}
\newcommand{\hSig}[1]{\textit{Fault signature#1}}

\title{\helium{}: Grouping Fuzzed Crashes Based On Fault Signatures}

\author{Ashwin Kallingal Joshy}
\orcid{0000-0001-8236-5027}
\affiliation{
  \institution{Iowa State University}
  \city{Ames}
  \state{Iowa}
  \country{USA}
}
\email{ashwinkj@iastate.edu}

\author{Wei Le}
\orcid{0000-0002-6797-0648}
\affiliation{
  \institution{Iowa State University}
  \city{Ames}
  \state{Iowa}
  \country{USA}
}
\email{weile@iastate.edu}

\copyrightyear{2022}
\acmYear{2022}
\setcopyright{rightsretained}
\acmConference[ASE '22]{37th IEEE/ACM International Conference on Automated Software Engineering}{October 10--14, 2022}{Rochester, MI, USA}
\acmBooktitle{37th IEEE/ACM International Conference on Automated Software Engineering (ASE '22), October 10--14, 2022, Rochester, MI, USA}
\acmDOI{10.1145/3551349.3556959}
\acmISBN{978-1-4503-9475-8/22/10}


\begin{document}
\begin{abstract}
Fuzzing has been an important approach for finding bugs and vulnerabilities in programs. Many fuzzers deployed in industry run daily and can generate an overwhelming number of crashes. Diagnosing such crashes can be very challenging and time-consuming. Existing fuzzers typically employ heuristics such as code coverage or call stack hashes to weed out duplicate reporting of bugs. While these heuristics are cheap, they are often imprecise and end up still reporting many ``unique'' crashes corresponding to the same bug. In this paper, we present \helium{} that uses \textit{fault signatures} to group crashes reported by the fuzzers. Fault signature is a small executable program and consists of a selection of necessary statements from the original program that can reproduce a bug. In our approach, we first generate a fault signature using a given crash. We then execute the fault signature with other crash inducing inputs. If the failure is reproduced, we classify the crashes into the group labeled with the fault signature; if not, we generate a new fault signature. After all the crash inducing inputs are classified, we further merge the fault signatures of the same root cause into a group. We implemented our approach in a tool called \helium{} and evaluated it on 3020 crashes generated from 15 real-world bugs and 4 large open source projects. Our evaluation shows that we are able to correctly group 99.1\% of the crashes and reported only 17 (+2) ``unique'' bugs, outperforming the state-of-the-art fuzzers.
\end{abstract}

\maketitle

\section{Introduction}\label{sec:Intro}
In the recent years, we have seen an increasing number of vulnerabilities in programs that were exploited~\cite{ProjectZero-2022,ZeroDayInitiative-2022}. Google's \textit{Project Zero}, a team of security researchers that study zero-day vulnerabilities, recorded their most ever detected and disclosed vulnerabilities in 2021~\cite{ProjectZero-2022}. 67\% of the actively exploited zero-day vulnerabilities in 2021, detected by \textit{Project Zero}, were from memory related bugs. Thankfully, the modern state-of-the-art (SOTA) fuzzers are adept at finding these types of bugs~\cite{Finding-bugs-in-SQLite,Ding_Goues_2021}. Thus, the companies such as Microsoft and Google invest heavily to develop and deploy effective fuzzers for daily uses. Open source platforms such as GitHub~\cite{ClusterFuzzLite} and GitLab~\cite{Coverage-guided-fuzz-gitlab} also provide friendly integration to run fuzzers. However, even with large scale fuzzing services, like Google's OSS-Fuzz~\cite{Serebryany_2017} and Microsoft's OneFuzz~\cite{OneFuzz_2020}, using fuzzers to find and fix bugs still involves considerable manual efforts~\cite{Chen_Groce_Zhang_Wong_Fern_Eide_Regehr_2013}. The fault diagnosis may hardly catch up to the speed at which new crashes are generated. As a result, critical and exploitable bugs can be left undiagnosed in the large number of crashes reported by the fuzzers.

In this paper, our goal is to develop approaches and tools that can help group fuzzed crashes and provide support for diagnosing them. Grouping fuzzed crashes is also called \textit{crash deduplication} or reporting \textit{unique crashes}~\cite{Klees_Ruef_Cooper_Wei_Hicks_2018}. In the past, the common approach for crash deduplication is to apply heuristic metrics~\cite{certBFF,swiecki2017honggfuzz,Cha_Woo_Brumley_2015,Rawat_Jain_Kumar_Cojocar_Giuffrida_Bos_2017,zalewski2017american,Gan_Zhang_Qin_Tu_Li_Pei_Chen_2018,Bohme_Pham_Nguyen_Roychoudhury_2017,van_Tonder_Kotheimer_Le_Goues_2018,Holmes_Groce_2018,Pham_Khurana_Roy_Roychoudhury_2017}, such as call stack hashing~\cite{certBFF,swiecki2017honggfuzz,Cha_Woo_Brumley_2015,Rawat_Jain_Kumar_Cojocar_Giuffrida_Bos_2017}, instrumented coverage information~\cite{zalewski2017american,Gan_Zhang_Qin_Tu_Li_Pei_Chen_2018,Bohme_Pham_Nguyen_Roychoudhury_2017} or dynamic symptoms~\cite{van_Tonder_Kotheimer_Le_Goues_2018,Holmes_Groce_2018,Pham_Khurana_Roy_Roychoudhury_2017} to compare the similarities of the crashes.  For example, AFL~\cite{zalewski2017american} uses instrumented branch coverage, while CERT-BFF~\cite{certBFF} and Honggfuzz~\cite{swiecki2017honggfuzz} use call stacks. However, the coverage based metrics tend to inflate the number of ``unique'' fuzzed crashes~\cite{Klees_Ruef_Cooper_Wei_Hicks_2018,Hazimeh_Herrera_Payer_2020} where typically crashes that traverse different paths will be separated, independent of whether they trigger the same bug. Meanwhile, call stack based metrics have risks of misclassifications; the crashes generated from the same bugs are separated into different groups as their call stacks are different~\cite{Blazytko_AURORA_2020} or different bugs are grouped together because they have the same call stacks~\cite{Pham_Khurana_Roy_Roychoudhury_2017}. The dynamic symptoms based approaches, like the ones based on symbolic analysis~\cite{Pham_Khurana_Roy_Roychoudhury_2017} and automatically generated patches~\cite{van_Tonder_Kotheimer_Le_Goues_2018,Holmes_Groce_2018}, are more precise, but they often have a limited scope, e.g., applicable only to a certain type of crashes.

In this paper, we propose to use \textit{fault signatures} to group the fuzzed crashes. A fault signature is an executable program that consists of  ``indispensable'' statements that can reproduce the bug. As opposed to call stacks, failure symptoms, and coverage based metrics, the fault signature captures the root cause of a failure.  Crashes grouped based on the same fault signature thus likely share the root cause and fix, and should be diagnosed together. We say these crashes are induced by the {\it same bug}.

Our approach consists of three components, namely \textit{generating fault signatures}, \textit{classify with fault signatures}, and \textit{merging fault signatures}. Given a collection of fuzzed crashes (here each crash is associated with an input), we first ran a crash inducing input to get its dynamic trace. We then create an executable program from the trace and perform program reduction to generate an as-small-as-possible program, namely \textit{fault signature}, that can reproduce the bug (meaning removal of any statement from this program can result in the bug not triggering).  Since the fault signature only contains a subset of statements from the complete crashing trace, the two crashes that exercise different paths in the original program can be grouped into the same signature as long as the two share the subset of root cause statements. To classify the next fuzzed crash, we took its crash inducing input and ran with the generated fault signatures and see if the failure has been produced with any of the fault signatures; if so, we classify the fuzzed crash into a group labeled with the fault signature. These two steps are implemented in the components of \textit{generating fault signatures} and \textit{classify with fault signatures} respectively.

After all the crashes have been bucketed into the groups, each of which is labeled with a fault signature, we perform post-processing, namely {\it merging fault signatures}, to examine if any of the two fault signatures can be further grouped. Two fault signatures may share a set of statements that are root causes, but differ in the paths that lead to the root cause from the entry of the program. These fault signatures can be grouped. We applied a heuristic based matching between two fault signatures to group very similar fault signatures. Specifically, we measure how many common statements the two signatures share and whether the call stacks are similar when crashing the two different fault signatures with their respective crash inducing inputs.

We implemented our approach in a tool called \helium{} for C programs. We used 15 real-world bugs from 4 large open source projects to evaluate it. Furthermore, we generated a total of 3020 fuzzed crashes and compared \helium{} with 3 SOTA fuzzers, name \textit{AFL}, \textit{CERT-BFF} and \textit{Honggfuzz}, and a total of 5 settings. We used developers' patches to establish the ground truth, similar to the approaches  in~\cite{Klees_Ruef_Cooper_Wei_Hicks_2018,van_Tonder_Kotheimer_Le_Goues_2018}. Our results show that we correctly group 2995 (99.1\%) of the 3020 fuzzed crashes and didn't misclassify any crashes into a wrong fault group. We grouped the fuzzed crashes from 15 different bugs into 17 (+2) groups while the next best baseline reported 40 groups, and other baselines reported 100+ or even 1000+ groups. Considering it is very time-consuming to diagnose a bug, our approach thus has a great potential to improve the productivity of bug diagnosis and fix associated with fuzzing tools.

In summary, we make the following contributions in the paper:
\begin{enumerate}
  \item we proposed to use \textit{fault signatures} to group fuzzed crashes, and our intuition is that fault signatures capture the root causes and thus can more accurately classify the crashes;
  \item we designed an algorithm, consisting of \textit{generating fault signatures}, \textit{classify with fault signatures}, and \textit{merging fault signatures}, to automatically group fuzzed crashes, and
  \item we implemented our tool \helium{} for C projects, and evaluated it with real-world bugs and large open source projects. Our results show that our approach can correctly group the crashes and significantly outperformed the SOTA widely deployed fuzzers.
\end{enumerate}

\section{Motivation}\label{sec:Motivation}
In this section, we provide a few  simple examples to explain the challenges of grouping fuzzed crashes and why existing fuzzing deduplication methods~\cite{certBFF,swiecki2017honggfuzz,Rawat_Jain_Kumar_Cojocar_Giuffrida_Bos_2017,zalewski2017american,Gan_Zhang_Qin_Tu_Li_Pei_Chen_2018,Bohme_Pham_Nguyen_Roychoudhury_2017} are not sufficient.

In Figure~\ref{subfig:CovDup}, we show a code snippet that contains a null-pointer dereference at line~3 adapted from~\cite{Klees_Ruef_Cooper_Wei_Hicks_2018}. Here, the pointer \texttt{p} at line 2 is not initialized, and then the dereference at the next line can lead to crash. This bug can be triggered independent of the condition outcome at line 8, as both the paths $\left<6,7,8,1,2,3\right>$ and $\left<6,7,9,1,2,3\right>$ calls into the \texttt{bug} function. Due to the presence of two distinct paths, a coverage based heuristic, e.g., used in AFL, will classify the fuzzed crashes for this bug into two separate groups. However, in this case, the branch at line~8 is not important for triggering the bug.

\begin{figure}[ht]
  \begin{subfigure}[b]{0.9\columnwidth}
    \centering
    \lstset{
      numbers=left,
      numbersep=5pt,
      belowcaptionskip=1\baselineskip,
      breaklines=true,
      xleftmargin=\parindent,
      language=C,
      showstringspaces=false,
      basicstyle=\footnotesize\ttfamily,
      keywordstyle={\bfseries\color{green!40!black}},
      commentstyle={\itshape\color{purple!40!black}},
      identifierstyle=\color{blue},
      stringstyle=\color{orange},
    }
\begin{lstlisting}
void bug() {
  int *p;
  int value = *p; // Null Pointer Deference
}
int main (int argc, char* argv[]) {
  if (argc >= 2) {
    char b = argv[1][0];
    if ( b == 'a' ) bug();
    else bug();
  }
  return 0;
}
\end{lstlisting}
    \caption{A null pointer dereference with different paths}\label{subfig:CovDup}
  \end{subfigure}

  \begin{subfigure}[b]{0.9\columnwidth}
    \centering
    \lstset{
      numbers=left,
      numbersep=5pt,
      belowcaptionskip=1\baselineskip,
      breaklines=true,
      xleftmargin=\parindent,
      language=C,
      showstringspaces=false,
      basicstyle=\footnotesize\ttfamily,
      keywordstyle={\bfseries\color{green!40!black}},
      commentstyle={\itshape\color{purple!40!black}},
      identifierstyle=\color{blue},
      stringstyle=\color{orange},
    }
\begin{lstlisting}
void bug() {
  int *p;
  int value = *p;
}
int main (int argc, char* argv[]) {
  if (argc >= 2) {
    bug();
  }
}
\end{lstlisting}
    \caption{\helium{}: \hSig{} for the bug in Figure~\ref{subfig:CovDup}}\label{subfig:CovDupSig}
  \end{subfigure}
  \caption{Deduplication based on code coverage can fail}\label{fig:SameBug}
  \Description{Examples to show deduplication based on code coverage can fail}
\end{figure}

In our approach, we will take a crashed input and collect its dynamic trace. We then reduce the trace to be able to reproduce the bug. We call such a program that contains only statements that contribute to the failure the {\it fault signature}. For Figure~\ref{subfig:CovDup}, we generate the fault signature shown in Figure~\ref{subfig:CovDupSig}. Here, lines 7--9 in the original program are replaced with a single call to the function {\tt bug}.  To group the fuzzed crashes, we run the crash inducing inputs with this fault signature. If the failure is reproduced with the fault signature, we group the fuzzed crash;  if not, we will generate a new fault signature that can represent the crash. In this example, we can group any crashes triggered along the branch at line 8 (inputs that start with {\tt a}) or along the one at line 9 (inputs that do not start with {\tt a}) into the same fault signature.

\begin{figure}
  \begin{subfigure}{.9\columnwidth}
    \lstset{
      numbers=left,
      numbersep=5pt,
      belowcaptionskip=1\baselineskip,
      breaklines=true,
      xleftmargin=\parindent,
      language=C,
      showstringspaces=false,
      basicstyle=\scriptsize\ttfamily,
      keywordstyle={\bfseries\color{green!40!black}},
      commentstyle={\itshape\color{purple!40!black}},
      identifierstyle=\color{blue},
      stringstyle=\color{orange},
    }
\begin{lstlisting}
#include<stdlib.h>
#include<stdbool.h>
void trigger(char *s) { int value = *s; }
void bug(char *s, bool b1) {
  if (b1) free(s); // Bug 1
  trigger(s);
}
void foo(char *s, bool b1) { bug(s, b1); }
void bar(char *s, bool b1) { bug(s, b1); }
int main (int argc, char* argv[]) {
  if (argc >= 2) {
    char *s = malloc(100);
    char b = argv[1][0];
    if ( b == 'a' ) foo(s, true);
    else bar(s, true);
  } else {
    char *p; // Bug 2
    foo(p, false);
  }
  return 0;
}
\end{lstlisting}
    \caption{\texttt{null pointer dereference} with different call stacks}\label{subfig:CallDup}
  \end{subfigure}
  \begin{subfigure}[b]{.7\columnwidth}
    \centering
    \lstset{
      belowcaptionskip=1\baselineskip,
      breaklines=true,
      xleftmargin=\parindent,
      language=C,
      showstringspaces=false,
      basicstyle=\footnotesize\ttfamily,
      keywordstyle={\bfseries\color{green!40!black}},
      commentstyle={\itshape\color{purple!40!black}},
      identifierstyle=\color{blue},
      stringstyle=\color{orange},
      morekeywords={Bug1,Bug2},
      escapechar=\!,
    }
\begin{lstlisting}
Bug1:
  Bug1-Crash1   main!$\rightarrow$\textcolor{olive}{foo}!!$\rightarrow$!bug!$\rightarrow$!trigger
  Bug1-Crash2   main!$\rightarrow$\textcolor{red}{bar}!!$\rightarrow$!bug!$\rightarrow$!trigger
Bug2:
  Bug2-Crash1   main!$\rightarrow$\textcolor{olive}{foo}!!$\rightarrow$!bug!$\rightarrow$!trigger

\end{lstlisting}
    \caption{call stacks for the bugs in Figure~\ref{subfig:CallDup}}\label{subfig:CallInfo}
  \end{subfigure}
  \begin{subfigure}[b]{.9\columnwidth}
    \centering
    \lstset{
      numbers=left,
      numbersep=5pt,
      belowcaptionskip=1\baselineskip,
      breaklines=true,
      xleftmargin=\parindent,
      language=C,
      showstringspaces=false,
      basicstyle=\scriptsize\ttfamily,
      keywordstyle={\bfseries\color{green!40!black}},
      commentstyle={\itshape\color{purple!40!black}},
      identifierstyle=\color{blue},
      stringstyle=\color{orange},
      morekeywords={Sig1,Sig2,Sig3},
      escapechar=\!,
    }
\begin{lstlisting}
Sig1 ===========
void trigger(char *s) { int value = *s; }
void bug(char *s, bool b1) {
  free(s); trigger(s);
}
void foo(char *s, bool b1) { bug(s, b1); }
int main (int argc, char* argv[]) {
  if (argc >= 2) {
    char *s = malloc(100);
    char b = argv[1][0];
    if ( b == 'a' ) !\colorbox{yellow!50}{foo(s, true);}! } }

Sig2 =============
void trigger(char *s) { int value = *s; }
void bug(char *s, bool b1) {
  free(s); trigger(s);
}
void bar(char *s, bool b1) { bug(s, b1); }
int main (int argc, char* argv[]) {
  if (argc >= 2) {
    char *s = malloc(100);
    char b = argv[1][0];
    if ( b == 'a' )!\colorbox{yellow!50}{;}!
    !\colorbox{yellow!50}{else bar}(s, true);! } }

Sig3 =============
void trigger(char *s) { int value = *s; }
void bug(char *s, bool b1) {
  trigger(s);
}
void foo(char *s, bool b1) { bug(s, b1); }
int main (int argc, char* argv[]) {
  if (argc >= 2) {
  } else {
    char *p;
    foo(p, false);
  }
}
\end{lstlisting}
    \caption{\helium{}: \hSig{s} for the bugs in Figure~\ref{subfig:CallDup}}\label{subfig:CallDupSig}
  \end{subfigure}
  \caption{Deduplication based on call stacks can fail}\label{fig:DiffBug}
  \Description{Example to show deduplication based on call stacks can fail}
\end{figure}

In the second example shown in Figure~\ref{subfig:CallDup}, we provide two different null pointer difference bugs.
\texttt{Bug1}, marked at line~5, will trigger an incorrect pointer dereference at line~3 after the pointer is freed at line~5. Meanwhile, \texttt{Bug2}, marked at line~17, will trigger a null pointer dereference at line~3 along the path $\left<17, 18, 8, 4,6,3\right>$. The uninitialized pointer at line~17 will be dereferenced.

The crashes for {\tt Bug1} can traverse different paths and lead to different call stacks at the crash site, as shown in Figure~\ref{subfig:CallInfo}. The first two lines indicate that \texttt{Bug1} can be triggered by calling \texttt{foo} (at line~14),  \texttt{bug} (at line~8), and \texttt{trigger} (at line~6); or by calling \texttt{bar} (at line~15),  \texttt{bug} (at line~9), and \texttt{trigger} (at line~6). Since the two crashes have two different call stacks, the call stacks based approach can fail to group them and will consider them as different bugs.

Similarly, the crash for {\tt Bug2} can be triggered by calling \texttt{foo} (at line~18),  \texttt{bug} (at line~8), and \texttt{trigger} (at line~6). As shown in Figure~\ref{subfig:CallInfo}, the call stacks for Bug1-Crash1 and Bug2-Crash1 are the same. Therefore, the call stack based deduplication methods can mistakenly group the two different bugs (they have different causes and require different fixes) into in one group.

Using our approach for this example, we will first generate fault signatures for the crashes, one at a time, shown in Figure~\ref{subfig:CallDupSig}, {\tt Sig1} at lines~2--11 for {\tt Bug1-Crash1},  {\tt Sig2} at lines~14--24 for {\tt Bug1-Crash2}, and {\tt Sig3} at lines~27--38 for {\tt Bug2-Crash1}. We can see that in the fault signatures, the statements that weren't contributing to the bug's manifestation have been removed.

Each fault signature can represent one path or a set of paths that lead the crashes. Since two sets of paths may contain the same root cause, we further merge fault signatures to be the same fault group. In this example, we observe that {\tt Sig1} and {\tt Sig2} differ only at the branch {\tt b==`a'} (highlighted in yellow and also see line~11 and lines 23--24), and we can merge them to be the same fault group. The merged fault signatures then can classify all fuzzed crashes that manifest \texttt{Bug1} into a single group. Similarly,  our approach will determine that {\tt Sig3} has no close relation with {\tt Sig1} and {\tt Sig2} in terms of branches and the shared statements. We thus classify it as a separate fault group.

The above examples show that our fault signature based grouping potentially is more accurate than coverage based and call stack based approaches that are popularly used in the current fuzzers. In Sections 3 and 4, we provide the details on how we generate fault signatures, how we group the fuzzed crashes based on the fault signatures, and how we further merge fault signatures into fault groups.

\section{Our approach}\label{sec:Overview}
\begin{figure*}
  \centering
  \def\svgwidth{0.9\textwidth}
  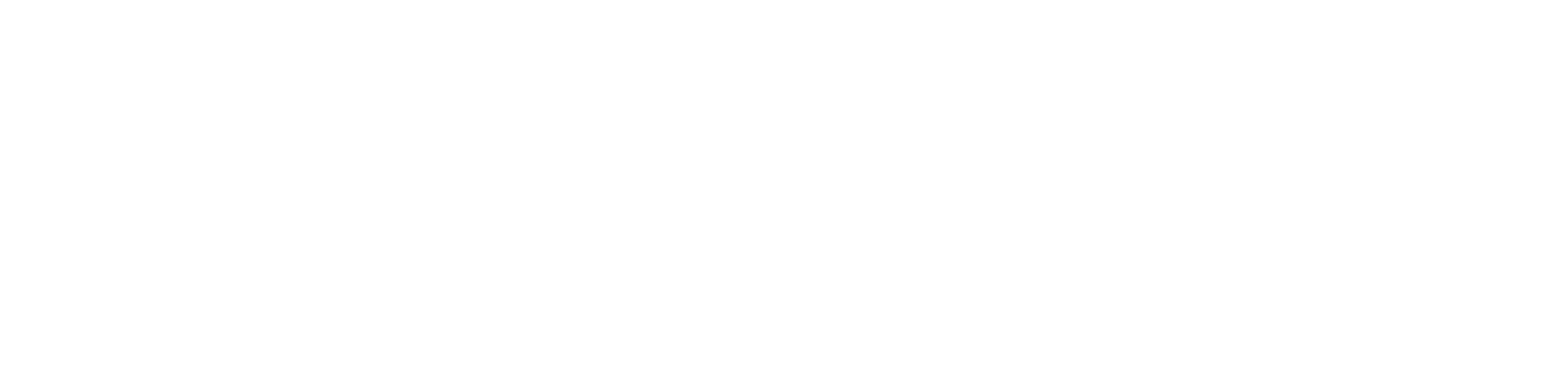
  \caption{Overview of \helium{}}\label{fig:Overview}
  \Description{Overview of \helium{} as a flow chart}
\end{figure*}

In this section, we first give an overview of \helium{}. Next, we provide a detailed explanation to help understand what is a \textit{fault signature}. We then present the three main components of \helium{}.

\subsection{An Overview}
Figure~\ref{fig:Overview} presents an overview of our workflow. \helium{} takes as input a collection of crashes from the fuzzers. Each crash is associated with an input. Each input will be run through a set of fault signatures created so far. If the failure is triggered with some fault signature, the fuzzed crash is put in a bucket labeled with the corresponding fault signature. This step is implemented in the component named \textit{Classify with Fault Signature}.

If the fuzzed crash input did not trigger any failures matching with existing fault signatures created so far, we run the input in the original program using PIN~\cite{Luk_Cohn_Muth_Patil_Klauser_Lowney_Wallace_Reddi_Hazelwood_2005} and collect its dynamic trace. We then patch the dynamic trace to generate an executable program. In the next step, we use a program reduction tool, C-Reduce, to reduce the executable program into the \textit{fault signature}. C-Reduce ensures that the fault signature can still reproduce the failure at the same location with the same failure symptoms while statements not relevant to the failure reproduction are removed.  This step is implemented in the component named \textit{Generate Fault Signatures}.

Once all the crashes and their inputs are labeled with the fault signatures, we perform the step of \textit{Merge Fault Signatures} and apply heuristics and similarity metrics to group fault signatures that likely originated from the same root cause. The resulting fault groups, representing grouped fuzzed crashes, and their corresponding fault signatures will be presented as the output of \helium{}.

\subsection{Fault Signatures}\label{subsec:FaultSig}
\begin{figure}
  \centering
  \lstset{
    numbers=left,
    numbersep=5pt,
    belowcaptionskip=1\baselineskip,
    breaklines=true,
    xleftmargin=10pt,
    language=C,
    showstringspaces=false,
    basicstyle=\footnotesize\ttfamily,
    keywordstyle={\bfseries\color{green!40!black}},
    commentstyle={\itshape\color{purple!40!black}},
    identifierstyle=\color{blue},
    stringstyle=\color{orange},
    escapechar=\!,
  }
\begin{lstlisting}
void HTMLlineproc0(!$\dots$!) {
  struct readbuffer *obuf = h_env->obuf;
  struct table_mode *tbl_mode = NULL;
  if (obuf->table_level >= 0) {
    int pre_mode = (obuf->table_level >= 0) ?
    tbl_mode->pre_mode : obuf->flag; // Null Pointer
  };
}
void loadHTMLstream(!$\dots$!) {
  !$\dots$!
  while ((lineBuf2 = StrmyUFgets(f))->length) {
    HTMLlineproc0(lineBuf2->ptr, &htmlenv1, internal);
  }
}
Buffer *loadHTMLBuffer(!$\dots$!) {
  !$\dots$!
  loadHTMLstream(f, newBuf, src, newBuf->bufferprop & BP_FRAME);
}
Buffer *loadGeneralFile(!$\dots$!) {
  !$\dots$!
  proc = loadHTMLBuffer;
}
int main(int argc, char **argv, char **envp) {
  !$\dots$!
  newbuf = loadGeneralFile(url, NULL, NO_REFERER, 0, request);
}
\end{lstlisting}
  \caption{Fault Signature for a \texttt{Null Pointer Dereference} in \texttt{w3m}}\label{fig:w3mSig1}
  \Description{Fault Signature for a \texttt{Null Pointer Dereference} in \texttt{w3m}}
\end{figure}

Fault signature can be viewed as a minimized version of the original program consisting of only the statements necessary for triggering a particular bug. Ideally, a fault signature can reproduce the same bug for all the inputs that can trigger the bug in the original program. The statements in a fault signature include two parts: (1) those that trigger the bug and (2) those that set up the necessary conditions, e.g., parsing the input, for reaching the buggy location.

As an example, consider the \texttt{null pointer dereference} bug~\footnote{https://github.com/tats/w3m/issues/18} in \texttt{w3m}\texttt{--0.5.3}. Even though the entire program consists of 80K lines of code, we generated a fault signature consisting of less than 100 lines. It can trigger the bug using the crash inducing inputs from the original programs. A simplified version of the fault signature is shown in Figure~\ref{fig:w3mSig1}. The \texttt{null pointer dereference} occurs at line~6, while trying to access the variable \texttt{tbl\_mode}. This variable was previously initialized as \texttt{NULL} at line 3 and not updated since then. Hence, lines 2--8 in Figure~\ref{fig:w3mSig1} can be considered as the statements that trigger the bug, and lines 9--26 are necessary to set up the conditions to reach the bug.

While the lines actually triggering the bug (lines 2--8 in Figure~\ref{fig:w3mSig1}) mostly remain the same between different crash inducing inputs, the statements leading to it (lines 9--26) can be different. In other words, a bug can be triggered when executing different paths (e.g., in the region of lines 9--26), but these paths can share a root cause (lines 2--8). As a concrete example, consider the two execution paths (\texttt{Bug1-Crash1} and \texttt{Bug1-Crash2} from Figure~\ref{subfig:CallInfo}) for the bug shown in Figure~\ref{subfig:CallDup}. The statements triggering this bug are in lines 3--7, while lines 8--15 provide the necessary conditions to reach the bug location.

It is difficult to enumerate all the different ways to reach a program point. Hence, producing an ideal fault signature that can reproduce the bug for all the crash inducing inputs is hard. However, since we have access to some inputs responsible for triggering the crashes, it is easy to create a fault signature based on one input with respect to its path, and then determine if other inputs can crash the same paths. Therefore, we chose to generate such fault signatures in \textit{Generate Fault Signature} to group the fuzzed crashes.

\subsection{Generate Fault Signatures}\label{subsec:GenFaultSig}
In order to generate fault signatures, we need to identify statements that are necessary for triggering a bug. Any statements that are not executed during a bug's manifestation are not necessary. So as a first step, we ran the crash inducing input with the original program to collect its dynamic trace. We used PIN, a dynamic binary instrumentation framework, to achieve this. The dynamic trace information collected using PIN is more resilient to call stack corruptions as opposed to the traditional methods~\cite{cert-bff-crash-recyler}.

The statements collected using dynamic trace typically don't include lines representing static information such as variables definitions, structure initialization and switch case headers. Or in other words, it is not possible to generate an executable fault signature directly using just the dynamic trace. Therefore, as the second step, we extracted all the functions that had a statement present in the dynamic trace. This takes care of missing local variable definitions and switch case headers. To get the global variable definitions and structure initialization, we extracted all global variables, macros, header file includes, and structure initialization using tools like srcML~\cite{Collard_Decker_Maletic_2013}. We made an executable program from these extracted information using the compilation and linker flags obtained via tools like Bear~\cite{bear-url}. This program, even though not minimal, is a reduced version of the original program capable of triggering the original bug.

\begin{figure}
  \centering
  \lstset{
    numbers=left,
    numbersep=5pt,
    belowcaptionskip=1\baselineskip,
    breaklines=true,
    xleftmargin=\parindent,
    language=C,
    showstringspaces=false,
    basicstyle=\footnotesize\ttfamily,
    keywordstyle={\bfseries\color{green!40!black}},
    commentstyle={\itshape\color{purple!40!black}},
    identifierstyle=\color{blue},
    stringstyle=\color{orange},
    escapechar=\!,
  }
\begin{lstlisting}
void HTMLlineproc0(!$\dots$!) {
  !\colorbox{red!50}{Lineprop mode;}!
  !\colorbox{red!50}{int cmd;}!
  struct readbuffer *obuf = h_env->obuf;
  !\colorbox{red!50}{int indent, delta;}!
  !\colorbox{red!50}{struct parsed\_tag *tag;}!
  !\colorbox{red!50}{Str tokbuf;}!
  struct table_mode *tbl_mode = NULL;
  !\colorbox{red!50}{if (w3m\_debug) \{}!
    !\colorbox{red!50}{HTMLlineproc1()}!
  !\colorbox{red!50}{\}}!
  !\colorbox{red!50}{tokbuf = Strnew();}!
  if (obuf->table_level >= 0) {
    !\colorbox{red!50}{char *str, *p;}!
    !\colorbox{red!50}{int is\_tag = FALSE;}!
    int pre_mode = (obuf->table_level >= 0) ?
    tbl_mode->pre_mode : obuf->flag; // Null Pointer
    !\colorbox{red!50}{// Rest of if block}!
  }
  !\colorbox{red!50}{else \{ // Else block \}}!
  !\colorbox{red!50}{// Rest of the function}!
}
!\colorbox{red!50}{void HTMLlineproc1 () \{$\dots$\}}!
!\colorbox{red!50}{str StrNew () \{$\dots$\}}!
\end{lstlisting}
  \caption{Example of the statements removed for creating the Fault Signature in Figure~\ref{fig:w3mSig1}}\label{fig:credRem}
  \Description{Example of the statements removed for creating the Fault Signature in Figure~\ref{fig:w3mSig1}}
\end{figure}

As the final step, we used C-Reduce~\cite{Regehr_Chen_Cuoq_Eide_Ellison_Yang_2012} to remove statements not required for triggering the bug to generate fault signatures. C-Reduce, by default, uses a set of 135 passes to minimize the program, which also include transformations such as renaming variables and function names and merging control branches. We developed a custom configuration of C-Reduce by removing 45 passes to suit our needs. Figure~\ref{fig:credRem} shows an example of the reduction (highlighted in red) when using our configuration for producing the fault signature shown in Figure~\ref{fig:w3mSig1}. Only the statements involving the variables \texttt{obuf} and \texttt{tbl\_mode} are necessary for reproducing the \texttt{null pointer dereference} bug at line 17. Hence, all the statements not related to the two variables till the fault's manifestation (at line 17) are removed (lines~2, 3, 5--7, 9--12, 14, 15) by C-Reduce. Since the statements after triggering the bug are also unnecessary, they also got removed (lines~18--21). We also remove the entire functions that were only used in the removed statements (lines~23 and 24), like \texttt{HTMLlineproc1} (used at line~10) and \texttt{StrNew} (used at line~12).

\subsection{Classify with Fault Signatures}\label{subsec:classifyWith}
Although the fault signature generation starts with one crash inducing input, after trace minimization and removing unnecessary statements, the fault signature can crash a set of failure inducing inputs that exercise the same path and the paths that only differ in the removed statements. It thus can be used to group a set of crash inducing inputs.

We ran crash inducing inputs with the existing fault signatures. If the same failure occurs (meaning it triggers the same bug type at the same source code location with the same call stack as the original input used to produce the fault signature), we classify the fuzzed crash into the group labeled with this fault signature. Otherwise, we take the input that can not yet be categorized and generate another fault signature.

When running a fuzzed crash with a fault signature, we may encounter failures that do not match the original crashes, e.g., entering an infinite loop or hanging indefinitely. Therefore, we set a configurable timeout (1 minute in our evaluation) when running fault signatures to classify whether a valid failure is triggered. We also encountered some flakiness when running with fault signatures caused by the nondeterminism in the software execution. Hence, we repeated running any fuzzed crash that failed to trigger a bug for a fixed number of times (10 times in our evaluation).

\subsection{Merge Fault Signatures}\label{subsection:GenFaultGroup}
Our fault signature is created from the dynamic trace generated using a single crash inducing input. As discussed in Section~\ref{subsec:GenFaultSig}, these fault signatures don't necessarily cover all the statements that can lead to the program state from which the bug can be triggered. Thus, during generation and classification of fault signatures, it is possible to create multiple fault signatures for the same bug, each of which represent a scenario of reaching the bug location. For example, see \texttt{Sig1} and \texttt{Sig2} for \texttt{Bug1} in Figure~\ref{fig:DiffBug}. In order to further group all the fuzzed crashes from \texttt{Bug1} into one group, we develop a technique to cluster fuzzed crashes associated with these fault signatures.

Our considerations are twofold. First, we want to group fault signatures of the same root cause (while a root cause can cover a segment/set of statements), and thus we should consider the similarity/overlap between the fault signatures. Second, we also consider the paths that lead to the crash site when merging the fault signatures. We observed that different bugs may fail at the same location, but two crashes that traverse very different paths before reaching the same location likely have different root causes. For example, in Figure~\ref{fig:DiffBug}, \texttt{Sig3} from \texttt{Bug2} shares the bug manifestation statements (line 3 in Figure~\ref{subfig:CallDup}) with \texttt{Sig1} and \texttt{Sig2} from \texttt{Bug1}. However, the actual root causes (line 17 for \texttt{Bug2} and line 5 for \texttt{Bug1} in Figure~\ref{subfig:CallDup}) along the paths leading to the buggy state (lines 18,4--7 for \texttt{Bug2} and lines 8--16 for \texttt{Bug1} respectively) are very different.

Specifically, to measure the similarity between two fault signatures, we used the Levenshtein's edit distance between them. See Equation~\ref{eq:SimFault}, where $Sim_{Sig}$ is the similarity score, $MAXSize$ returns the maximum size in lines of code of the two fault signatures, $S_1$ and $S_2$, and $LDistance$ is the Levenshtein's edit distance between the two signatures.
\begin{equation}\label{eq:SimFault}
  Sim_{Sig} = \frac{MAXSize(S_1, S_2) - LDistance(S_1, S_2)}{MAXSize(S_1, S_2)}
\end{equation}

We also measured the similarity in the paths leading to the failure location using call stacks generated by running the crash inducing inputs with the fault signatures. Specifically, we used Equation~\ref{eq:SimCall}, where $Sim_{Call}$ is the similarity score, $COMMON$ is the number of functions that are shared between two call stacks $CS_1$ and $CS_2$, and $MAXSize$ is the maximum size in count of call stacks.
\begin{equation}\label{eq:SimCall}
  Sim_{Call} = \frac{COMMON(CS_1, CS_2)}{MAXSize(CS_1, CS_2)}
\end{equation}

The final similarity score used to decide whether two fault signatures should be merged or not is shown in Equation~\ref{eq:SimScore}, which is the average of the two similarity scores.
\begin{equation}\label{eq:SimScore}
  Sim_{Score} = \frac{Sim_{Sig} + Sim_{Call}}{2}
\end{equation}

\section{Put together: the algorithm}\label{sec:Approach}
In Algorithm~\ref{alg:Grouping}, we present our algorithm for grouping fuzzed crashes. The algorithm takes as input $C_I$, a collection of inputs that can lead to the crashes in a fuzzer, and generates the fault groups, each of which represents a bug and has the corresponding fault signature(s). The crashes can be generated from different runs of the same fuzzer or even from different fuzzers.

\begin{algorithm}
  \caption{Grouping fuzzed crashed using fault signatures}\label{alg:Grouping}
  \begin{algorithmic}[1]
    \STATE{\textbf{INPUT}: $C_I$ (Fuzzed crashes)}
    \STATE{\textbf{OUTPUT}: $F_g$ (Fault groups), $F_s$ (Fault signatures)}
    \STATE{Initialize $F_s$} \COMMENT{Fault signatures}\label{AlgLine:initFS}
    \STATE{Initialize $F_g$} \COMMENT{Fault groups}\label{AlgLine:initFG}

    \FORALL{$c \in C_I$}\label{AlgLine:forFS}
      \IF{$\exists s \in F_s \text{ and } s \rightarrow c$}\label{AlgLine:ifExistFS} \STATE{Add $c$ to  $s.Crashes$}\label{AlgLine:updateS}
      \ELSE{}\label{AlgLine:elseExistFS}
        \STATE{$t \leftarrow$ Dynamic Trace ($c$)}\label{AlgLine:dynTrace}
        \STATE{$s_{new} \leftarrow$ Generate Signature ($t$)}\label{AlgLine:genSig}
        \STATE{Add $s_{new}$ to $F_s$}\label{AlgLine:addNewtoFS}
      \ENDIF{}\label{AlgLine:endExistFS}
    \ENDFOR{}\label{AlgLine:endforFS}

    \STATE{$worklist \leftarrow F_s$}\label{AlgLine:worklistInit}
    \WHILE[Merge Fault Signatures]{$worklist \neq \emptyset$}\label{AlgLine:whileWork}
      \STATE{Remove $s_c$ from $worklist$}\label{AlgLine:getfromWL}
      \STATE{Initialize $G_{sc}$} \COMMENT{Fault group for $s_c$}\label{AlgLine:initGsc}
      \FORALL{$s_{wl} \in worklist$}\label{AlgLine:forSwlInWL}
        \STATE{$score \leftarrow$ Compute Similarity ($s_c,s_{wl}$)}\label{AlgLine:compSim}
        \IF{$score \geq$ threshold}\label{AlgLine:checkThres}
        \STATE{Add $s_{wl}$ to $G_{sc}$}\label{AlgLine:addToGsc}
        \ENDIF{}\label{AlgLine:endCheckTres}
      \ENDFOR{}\label{AlgLine:endForSWL}
    \ENDWHILE{}\label{AlgLine:endWhileWork}
    \RETURN{$F_g$}\label{AlgLine:retuFg}
  \end{algorithmic}
\end{algorithm}

Lines~3--13 implements the components of \textit{Generate fault signatures} and \textit{Classify fault signatures} specified in Figure~\ref{fig:Overview}. Lines 14--25 implements the \textit{Merge fault signatures}. Specifically, at lines~\ref{AlgLine:initFS} and~\ref{AlgLine:initFG}, we initialize fault signatures ($F_s$) and fault groups ($F_g$) respectively. The initialization can either set them as empty sets or using existing fault signatures and fault groups from previous fuzzing campaigns or a different fuzzer. Lines~\ref{AlgLine:forFS}--\ref{AlgLine:endforFS} loop through all the fuzzed crashes ($C_I$) given as the inputs to create and test with fault signatures. At line~\ref{AlgLine:ifExistFS}, the current fuzzed crash ($c$) is checked against all the existing fault signatures to see if there exist a fault signature ($s$) that can lead it to crash. In case such an existing fault signature is found, at line~\ref{AlgLine:updateS}, we add the crash into the group represented by the fault signature. If we are unable to find such a fault signature, we create a new one at lines~\ref{AlgLine:elseExistFS}--\ref{AlgLine:endExistFS}. To create a fault signature, we run the program with the crashing input to collect its dynamic trace ($t$) at line~\ref{AlgLine:dynTrace}. This trace is used to create a new fault signature at line~\ref{AlgLine:genSig}, by removing the statements in the trace until the failure cannot be reproduced. The newly created signature ($s_{new}$) is added to other fault signatures at line~\ref{AlgLine:addNewtoFS}.

Once we have generated the fault signatures that can classify all the fuzzed crashes, we further merge the fault signatures into fault groups at lines~\ref{AlgLine:whileWork}--\ref{AlgLine:endWhileWork}. For this, we first create a work list ($worklist$) consisting of all the fault signatures at line~\ref{AlgLine:worklistInit}. We initialize a fault group ($GS_{sc}$) at line~\ref{AlgLine:forSwlInWL} for $s_c$. This can either be empty or be an existing fault group given at line~4.  We then traverse the work list at  lines~\ref{AlgLine:forSwlInWL}--\ref{AlgLine:endForSWL}. For each fault signature in the work list, we compute similarity scores ($score$) at line~\ref{AlgLine:compSim} and compare it with the other fault signatures in the work list. The $score$ is computed as the average of (1) normalized Levenshtein's edit distance and (2) normalized percent of matching functions in the crashing call stack between two fault signatures. If this similarity score is above a set threshold, then we add that fault signature ($s_{wl}$) to the fault group ($G_{sc}$) at line~\ref{AlgLine:addToGsc}.

Once the \texttt{worklist} is empty, we finish grouping all the fault signatures into fault groups. We return this group of fault groups, that represent ``unique'' crashes from the fuzzed crashes, as the output of our algorithm at line~\ref{AlgLine:retuFg}.

\section{Evaluation}\label{sec:Eval}
Our evaluation aims to answer two research questions:
\begin{itemize}
  \item \textbf{RQ1:} Can we correctly group crashes generated by the fuzzers?
  \item \textbf{RQ2:} How effective is our technique compared to the SOTA\@ methods?
\end{itemize}

\subsection{Experimental Setup}\label{subsec:ExpSetup}
\subsubsection{Implementation}
We implemented \helium{} for C programs using Clang~\cite{Lattner_Adve_2004}, srcML~\cite{Collard_Decker_Maletic_2013}, SQLite~\cite{sqlite3}, PIN~\cite{Luk_Cohn_Muth_Patil_Klauser_Lowney_Wallace_Reddi_Hazelwood_2005}, C-Reduce~\cite{Regehr_Chen_Cuoq_Eide_Ellison_Yang_2012} and Rust~\cite{Matsakis_Klock_2014}. Specifically, we used Clang, srcML and SQLite to create a database containing the function names and their line numbers at a file level granularity for each benchmark. Then we used this database with PIN to collect statement level dynamic traces for crash inducing inputs and generated an executable program. We used C-Reduce to minimize the executable programs into fault signatures. We used Rust to implement the similarity comparison between traces and between fault signatures. We used a cost of 1 for all the edits when computing the Levenshtein's distance. The threshold for grouping two \hSig{s} (line~\ref{AlgLine:checkThres} in Algorithm~\ref{alg:Grouping}) was set as ``0.7''.

\subsubsection{Subject Selection} To answer the research questions and demonstrate that our techniques are applicable in practice, we aim to use the benchmarks that (1) are real-world open-source C programs, (2) preferably contain multiple real-world bugs in each program, so that we can evaluate if our approach can separate the crashes from different bugs, (3) the bugs and their patches are known, so we can have a ground truth to compare against, (4) the bugs can be triggered by the fuzzers, so we can generate the crash corpus, and (5) can be handled by our baseline methods, so we can compare with them.

We searched for the readily available benchmarks based on the above 5 criteria in fuzzing literature~\cite{Pham_Khurana_Roy_Roychoudhury_2017,Gan_Zhang_Qin_Tu_Li_Pei_Chen_2018,Bohme_Pham_Nguyen_Roychoudhury_2017,Klees_Ruef_Cooper_Wei_Hicks_2018,Boehme_Cadar_Roychoudhury_2021,Liang_Pei_Jia_Shen_Zhang_2018,van_Tonder_Kotheimer_Le_Goues_2018}. We also went through the list of programs at AFL's website~\cite{afl-trophy}. As a result, we collected all the C projects (3 out of 6 total projects) provided by~\cite{van_Tonder_Kotheimer_Le_Goues_2018}, namely \texttt{w3m}, \texttt{sqlite} and \texttt{libmad}, and \texttt{libarchive} from AFL's website. We were not able to use the other 3 benchmarks from~\cite{van_Tonder_Kotheimer_Le_Goues_2018}, namely \texttt{PHP}, \texttt{R} and \texttt{Conntrackd}, as they were either implemented using multiple languages (PHP, R) or the bug/patch were in a non C file. Similarly, for C only benchmarks with multiple bugs on the AFL's websites, such as \textit{audiofile} and \texttt{libxml}, we were either not able to get the crashing input and the minimal patch, or they had the bug/patch in non C files.

Through the above process, we collected a total of 15 known bugs from 4 large real-world C projects. Specifically, we used 4 bugs from \texttt{w3m}, a text-based web browser, 8 bugs from \texttt{sqlite}, a database software, 1 bug from \texttt{libmad}, an MPEG audio decoding library, and 2 bugs from \texttt{libarchive}, a multi-format archive and compression library. Four \texttt{sqlite} bugs from~\cite{van_Tonder_Kotheimer_Le_Goues_2018} were not included due to the problems of reproducing them with PIN\@.  The projects, their sizes\footnote{LOC calculated using \texttt{tokei}, https://github.com/XAMPPRocky/tokei}, and the bugs are listed in the first three columns in Table 1.

\subsubsection{Fuzzer selection}
To demonstrate the effectiveness of our techniques and compare with meaningful baselines, we looked for fuzzers that (1) are open source, (2) are well documented, (3) have been widely used in research or industrial settings, (4) applied different methodologies for deduplicating fuzzed crashes (so we can compare with different approaches of deduplication used in practice), and (5) could work with our benchmarks. In case of fuzzers with similar deduplication methodologies, we picked the one that reported more bugs, cited by more references, and that are easier to work with our benchmarks. In the end, we used AFL~\cite{zalewski2017american} to generate crashes, and used the deduplication methods implemented in AFL, CERT-BFF~\cite{certBFF} and Honggfuzz~\cite{swiecki2017honggfuzz} as our baselines. Specifically, AFL uses branch (edge) coverage and coarse grained branch-taken hit counter to determine unique fuzzed crashes. Only fuzzed crashes associated with execution paths that involves new edges or doesn't visit the common edges are kept after deduplication. CERT-BFF uses hashes generated from last $N$ calls (frames) in the call stack to determine uniqueness. Any fuzzed crashes sharing the same call stack hash is discarded during the deduplication process. Honggfuzz, on the other hand, uses the information at the crash location (fault address, last known PC instruction and last 7 frames in call stack) to deduplicate the fuzzed crashes.

\begin{table*}[ht]
\begin{tabular}{@{}lccccccccc@{}}
\cmidrule(r){1-9}
\textbf{Benchmark} &
\multicolumn{1}{l}{\textbf{Size (KLOC)}} &
\multicolumn{1}{l}{\textbf{Bug ID}} &
\multicolumn{1}{l}{\textbf{Crashes}} &
\multicolumn{1}{l}{\textbf{Fault Sig}} &
\multicolumn{1}{l}{\textbf{Group}} &
\multicolumn{1}{l}{\textbf{Correct}} &
\multicolumn{1}{l}{\textbf{Incorrect}} &
\multicolumn{1}{l}{\textbf{Missed}} \\ \cmidrule(r){1-9}
\textbf{w3m}                 & \textbf{80.4}  & 1           & 250           & 1           & 1                        & 250           & 0          & 0            \\
v0.5.3                       &                & 2           & 352           & 4           & 1                        & 351           & 0          & 1            \\
                             &                & 3           & 250           & 4           & 1                        & 250           & 0          & 0            \\
                             &                & 4           & 139           & 2           & 1                        & 115           & 0          & 24           \\ \cmidrule(r){1-9}
Sub Total                    &                & \textbf{4}  & \textbf{991}  & \textbf{11} & \textbf{4}               & \textbf{966}  & \textbf{0} & \textbf{25}  \\ \cmidrule(r){1-9}
\textbf{SQLite}              & \textbf{313.3} & 5           & 285           & 5           & 1                        & 285           & 0          & 0            \\
v3.8.5                       &                & 6           & 191           & 14          & 1                        & 191           & 0          & 0            \\
                             &                & 7           & 240           & 5           & 1                        & 240           & 0          & 0            \\
                             &                & 8           & 113           & 2           & 1                        & 113           & 0          & 0            \\
                             &                & 9           & 226           & 2           & 1                        & 226           & 0          & 0            \\
                             &                & 10          & 237           & 5           & {\color[HTML]{CC0000} 2} & 237           & 0          & 0            \\
                             &                & 11          & 250           & 4           & 1                        & 250           & 0          & 0            \\
                             &                & 12          & 236           & 3           & {\color[HTML]{CC0000} 2} & 236           & 0          & 0            \\ \cmidrule(r){1-9}
Sub Total                    &                & \textbf{8}  & \textbf{1778} & \textbf{40} & \textbf{10}              & \textbf{1778} & \textbf{0} & \textbf{0}   \\ \cmidrule(r){1-9}
\textbf{libmad} v0.15.1b     & \textbf{18.0}  & 13          & 99            & 2           & 1                        & 99            & 0          & 0            \\ \cmidrule(r){1-9}
Sub Total                    &                & \textbf{1}  & \textbf{99}   & \textbf{2}  & \textbf{1}               & \textbf{99}   & \textbf{0} & \textbf{0}   \\ \cmidrule(r){1-9}
\textbf{libarchive}          & \textbf{207.2} & 14          & 67            & 1           & 1                        & 67            & 0          & 0            \\
v3.1.0                       &                & 15          & 85            & 1           & 1                        & 85            & 0          & 0            \\ \cmidrule(r){1-9}
Sub Total                    &                & \textbf{2}  & \textbf{152}  & \textbf{2}  & \textbf{2}               & \textbf{152}  & \textbf{0} & \textbf{0}   \\ \cmidrule(r){1-9}
\textbf{Total}               & \textbf{618.9} & \textbf{15} & \textbf{3020} & \textbf{55} & \textbf{17}              & \textbf{2995} & \textbf{0} & \textbf{25}  \\ \cmidrule(r){1-9}
\end{tabular}
\caption{Result of RQ1: Evaluating Grouping Correctness}\label{tab:RQ1}
\end{table*}

\subsubsection{Experimental design for RQ1}\label{subsubsec:GroupCorrect}
In RQ1, our goal is to evaluate the correctness of the grouping made by \helium{}. Specifically, we aim to discover (1) if we can correctly group all the fuzzed crashes from the same bug, (2) if we would incorrectly mix crashes from different bugs and put them in one group, and (3) if we would fail to group any fuzzed crashes.

To establish the ground truth, we propose to generate the crashes for the multiple known bugs and see if we can group crashes caused by the same bug and separate the crashes generated from different bugs. The challenge we face is that given an arbitrary given seed, the fuzzers may not trigger the known bugs or trigger them within a reasonable time window. To set up this experiment, we used a special configuration of the fuzzers together with the bug patches from the developers to achieve the goal. Specifically,  our approach is to run
AFL in the ``Crash Exploration Mode~\cite{afl-crash-mode}''. It takes a known fuzzed crash inducing input and uses the traditional feedback and genetic algorithms to quickly generate a corpus of crashes that explore different paths that can lead to similar crash state. We found that this approach is likely to generate crashes that trigger the given bug.  In our experiments, we found only a total of 28 among thousands of crashes that belong to some unknown bugs.

Our setup is as follows. First, we generate crash corpus for individual know bugs using the above approach. We ran AFL for 2 hours per bug. Given 2 hours, some bug generated more than 2K fuzzed crashes, some bug only generated less than 100 crashes. To balance the crashes from different bugs, we used at most 250 crashes from each bug (randomly select 250 if there are more than 250) and mixed them as a mixed crash corpus. Since this approach is heuristic, we also used the developers' patch to further validate whether the generated crashes are indeed from a known bug and which known bug it belongs to. Using developers' patch to determine ground truth~\cite{Klees_Ruef_Cooper_Wei_Hicks_2018,van_Tonder_Kotheimer_Le_Goues_2018}, is based on the assumption that if an input $I$ crashes a program $P$, but no longer crashes it after applying patch $p$, we can associate $I$ with the bug for $p$~\cite{Chen_Groce_Zhang_Wong_Fern_Eide_Regehr_2013}. Further, if two inputs $I_1$ and $I_2$ both crash $P$, but disappear with patch $p$, then both $I_1$ and $I_2$ are caused by the same bug (given that the patches are ``minimal''~\cite{WhatIsBug_2015}). As a result of validation, each crash is labeled with a bug ID and the ones do not match any existing bugs are labeled as unknown. We then apply \helium{} to group these crashes that we know the ground truth.

The validation with developers' patch is done on the mixed crash corpus consisting of up to 250 randomly selected fuzzed crash from each known bug. Due to the nature of fuzzing, it is possible for the fuzzing campaign to expose a different bug than the seed bug. For example, the fuzzing campaign for \textit{Bug 4} from \texttt{w3m} also generated crashes for \textit{Bug 2} which got selected during the random selection. However, after the validation these fuzzed crashes are correctly labeled with bug ID for \textit{Bug 2}. Hence, some bugs (\textit{Bug 2} and \textit{Bug 5} in Table~\ref{tab:RQ1} and Table~\ref{tab:RQ2}) have more than 250 fuzzed crashes (under \textit{Crashes}) associated with them.

To evaluate the correctness, we used as metrics of the number of fuzzed crashes that were (1) correctly grouped with fault groups for a bug, (2) incorrectly grouped with fault groups from a different bug, (3) weren't grouped to any fault group. We also recorded the number of fault signatures and fault groups created for each bug to measure the usefulness of the grouping.

\subsubsection{Experimental design for RQ2}\label{subsubsec:EvalFuzzer}
In RQ2, we compared \helium{} with the three SOTA real-world fuzzers regarding their capabilities of deduplicating crashes. In the following, we present the setups of the fuzzers used in comparison:

For AFL, we ran \textit{afl-cmin} on all the fuzzed crashes generated for each benchmark. It finds the smallest subset of fuzzed crashes that still exercises the full range of instrumented data points as the original fuzzed crash corpus. The remaining fuzzed crashes are reported as the deduplicated fuzzed crashes for \textit{AFL}.

We ran \textit{CERT-BFF} on all the fuzzed crashes for each benchmark, with \texttt{backtracelevels} set to 5 (\textit{BFF-5}). This gives us deduplicated fuzzed crashes based on the uniqueness of the last 5 frames (function calls) on the stack. Similarly, we also performed deduplication using the last frame (crashing function) of the call stack by setting \texttt{backtracelevels} to 1 (\textit{BFF-1}). We chose these two configurations because \textit{BFF-5} represents the default deduplication used by \textit{CERT-BFF}, and \textit{BFF-1} is used as a baseline in the related work~\cite{van_Tonder_Kotheimer_Le_Goues_2018}.

For \textit{Honggfuzz}, we ran the fuzzed crashes for each benchmarks with the \texttt{instrument} option enabled. This gave us deduplicated fuzzed crashes determined using a combination of code coverage, call stack, and crash site information~\cite{honggfuzz-usage}. Then we used the \texttt{noinst} mode (\textit{Honggfuzz-S}) to obtain deduplicated fuzzed crashes determined using only call stack and crash site information. We chose these two configurations because \textit{Honggfuzz} represents the default deduplication of the fuzzer and \textit{Honggfuzz-S} is also used as a baseline in the related work~\cite{van_Tonder_Kotheimer_Le_Goues_2018}.

In the experiment, we first collect a set of crashes for a project, e.g., 991 for w3m shown in Table 2. We then run a baseline tool, e.g., AFL, to deduplicate the crashes. The number of groups reported by the tool is listed under the columns of each baseline's \texttt{SubTotal} row, e.g., 490 for AFL in Table~\ref{tab:RQ2}. We then use the developer's patch to determine how many groups were reported for each bug, e.g., 109 for Bug 1 for AFL\@.

\subsubsection{Running the experiments} The initial crash corpus generation and the crash deduplication for the baseline fuzzers were run on a VM with 64-bit 32 core Intel Haswell processors. The \helium{} experiments were conducted on a VM with 64-bit 12 core Intel Haswell processors. Both the VMs had with 32 GB memory and were running CentOS 8.

\subsection{Results for RQ1: Grouping Correctness}\label{subsec:ResultCorrect}
\begin{table*}[ht]
\begin{tabular}{@{}lccccccccc@{}}
\cmidrule(r){1-9}
\textbf{Benchmark} &
  \multicolumn{1}{l}{\textbf{Bug ID}} &
  \multicolumn{1}{l}{\textbf{Crashes}} &
  \multicolumn{1}{l}{\textbf{FuzzerAid}} &
  \multicolumn{1}{l}{\textbf{AFL}} &
  \multicolumn{1}{l}{\textbf{BFF-5}} &
  \multicolumn{1}{l}{\textbf{BFF-1}} &
  \multicolumn{1}{l}{\textbf{Honggfuzz}} &
  \multicolumn{1}{l}{\textbf{Honggfuzz-S}} \\ \cmidrule(r){1-9}
\textbf{w3m}         & 1           & 250           & 1           & 109           & 2                        & 1                        & 49           & 49           \\
                     & 2           & 352           & 1           & 208           & 2                        & 1                        & 9            & 8            \\
                     & 3           & 250           & 1           & 113           & 89                       & 7                        & 14           & 17           \\
                     & 4           & 139           & 1           & 60            & 4                        & 3                        & 8            & 7            \\ \cmidrule(r){1-9}
Sub Total            & \textbf{4}  & \textbf{991}  & \textbf{4}  & \textbf{490}  & \textbf{97}              & \textbf{12}              & \textbf{80}  & \textbf{81}  \\ \cmidrule(r){1-9}
\textbf{SQLite}      & 5           & 285           & 1           & 179           & 14                       & 5                        & 2            & 3            \\
                     & 6           & 191           & 1           & 43            & 22                       & 8                        & 11           & 12           \\
                     & 7           & 240           & 1           & 81            & 7                        & 4                        & 1            & 1            \\
                     & 8           & 113           & 1           & 43            & 2                        & 3                        & 1            & 1            \\
                     & 9           & 226           & 1           & 60            & 3                        & 1                        & 1            & 1            \\
                     & 10          & 237           & 2           & 58            & 4                        & 1                        & 2            & 2            \\
                     & 11          & 250           & 1           & 134           & 1                        & 1                        & 2            & 2            \\
                     & 12          & 236           & 2           & 62            & 4                        & 2                        & 2            & 2            \\ \cmidrule(r){1-9}
Sub Total            & \textbf{8}  & \textbf{1778} & \textbf{10} & \textbf{660}  & \textbf{57}              & \textbf{25}              & \textbf{22}  & \textbf{24}  \\ \cmidrule(r){1-9}
\textbf{libmad}      & 13          & 99            & 1           & 58            & 4                        & 2                        & 5            & 6            \\ \cmidrule(r){1-9}
Sub Total            & \textbf{1}  & \textbf{99}   & \textbf{1}  & \textbf{58}   & \textbf{4}               & \textbf{2}               & \textbf{5}   & \textbf{6}   \\ \cmidrule(r){1-9}
\textbf{libarchive}  & 14          & 67            & 1           & 24            & {\color[HTML]{CC0000} 0} & {\color[HTML]{CC0000} 0} & 1            & 1            \\
                     & 15          & 85            & 1           & 44            & {\color[HTML]{CC0000} 1}                        & {\color[HTML]{CC0000} 1}                       & 1            & 1            \\ \cmidrule(r){1-9}
Sub Total            & \textbf{2}  & \textbf{152}  & \textbf{2}  & \textbf{68}   & \textbf{1}               & \textbf{1}               & \textbf{2}   & \textbf{2}   \\ \cmidrule(r){1-9}
\textbf{Total}       & \textbf{15} & \textbf{3020} & \textbf{17} & \textbf{1276} & \textbf{159}             & \textbf{40}              & \textbf{109} & \textbf{113} \\ \cmidrule(r){1-9}
\end{tabular}
\caption{Results of RQ2: Comparing \helium{} against SOTA fuzzer deduplication}\label{tab:RQ2}
\end{table*}

Table~\ref{tab:RQ1} shows the result for RQ1. Each row corresponds to a known bug labeled with \textit{Bug ID}. The column \textit{Crashes} lists the number of fuzzed crashes generated for each bug using the approach presented in  Section~\ref{subsubsec:GroupCorrect}.  The crashes reported in this column are post-processed using the developer's patch. The \textit{Fault Sig} and \textit{Group} columns provide the number of fault signatures and the number of fault groups generated for the known bug using \helium{}. Under \textit{Correct} column we list the number of fuzzed crashes that were correctly grouped in one of the fault groups generated for the bug. Similarly, the \textit{Incorrect} column lists the number of fuzzed crashes from unrelated bugs that were incorrectly grouped under one of the fault groups generated for the bug. Any fuzzed crash that \helium{} failed to group under any fault group is reported under \textit{Missed}.

Our results indicate that among the total 3020 fuzzed crashes for which we know the ground truth,  \helium{} correctly grouped 2995 (99.1\%) fuzzed crashes. For 3 benchmarks (\texttt{sqlite}, \texttt{libmad} and \texttt{libarchive}), we correctly classified 100\% (2029) of their fuzzed crashes. While we were unable to classify 25 (0.08\%) fuzzed crashes, we didn't misclassify any fuzzed crash into unrelated fault groups. The 25 fuzzed crashes we missed for \texttt{w3m} (1 from \textit{Bug 2} and 24 from \textit{Bug 4}) were due to fault signature generation failure caused by the project specific hard-coded dynamic functions. This implementation issue of \helium{} could be improved in the future.

\helium{} generated a total of 17 fault groups for the 15 known bugs. For 3 benchmarks (\texttt{w3m}, \texttt{libmad} and \texttt{libarchive}), we reported the same number of groups as the ground truth. The 2 extra fault groups (highlighted in red) generated for \texttt{sqlite} (one for \textit{Bug 10} and  one for \textit{Bug 12}) missed the clustering threshold by a very small margin (difference of 0.08\% and 0.4\% respectively).

We reported a total of 55 fault signatures for the 15 bugs sized between 52 LOC to 340 LOC\@. The number of fault signatures can indicate the number of different important paths (or scenario) in which a particular bug can manifest. Of particular interest is \textit{Bug 6} from \texttt{sqlite}, which produced 14 fault signatures from \textit{just} 191 fuzzed crashes. The relatively high number of fault signatures may be an indicator that this bug can be crashed from a variety of scenarios and thus likely more important.

Using AFL ``Crash Exploration Mode'', our crash corpus also included 28 fuzzed crashes that do not belong to any known bugs which we discovered using the developers' patches (See Section 5.1.4). \helium{} is able to successfully separate them into different groups from the known bugs.

\subsection{Results for RQ2: Comparing Against SOTA}\label{subsec:ResultCompare}
Table~\ref{tab:RQ2} shows the result for RQ2. Similar to Table~\ref{tab:RQ1}, each row corresponds to a known bug. We label them using the same assigned \textit{Bug ID} in Table~\ref{tab:RQ1}. For all the crashes listed under \textit{Crashes}, the grouping results from \helium{} and our baselines are listed under \textit{FuzzerAid}, \textit{AFL}, \textit{BFF-5}, \textit{BFF-1}, \textit{Honggfuzz} and \textit{Honggfuzz-S} respectively.

Our results show that \helium{}'s grouping is the same as the ground truth, except that we generated two additional groups for the \textit{SQLite} bugs. We generated a total of 17 groups for 15 bugs, compared to 40 from \textit{BFF-1}, 109 from \textit{Honggfuzz}, 113 from \textit{Honggfuzz-S}, 159 from \textit{BFF-5}, and 1276 from \textit{AFL}. Considering the challenges and cost of diagnosing a bug, our precise grouping techniques and improvement over the baselines indeed have practical values.

AFL used a conservative approach and applied branch coverage information to group the crashes of the same paths, thus it generated the most group. On the other hand, call stack hashes based approach of CERT-BFF and Honggfuzz were able to greatly reduce the number of groups, but with the risk of misclassification. For example, when we inspected the correctness of the grouping, we found that both \textit{BFF-1} and \textit{BFF-5} incorrectly classified all the fuzzed crashes from the two bugs of \texttt{libarchive} into one single group. See the numbers in the row of \textit{libarchive} highlighted in red.

\subsection{Summary}
In our evaluation, \helium{} is able to correctly group 2995 out of 3020 (99.1\%) fuzzed crashes without any incorrect classification. We were also the closest to ground truth in terms of grouping with 17 fault groups reported instead of ground truth's 15. The next closest baseline (BFF-1) reported 40 groups (2.35 times more) while still misclassifying fuzzed crashes from one group for \texttt{libarchive}.

The trace generated by PIN for creating the fault signatures varied between 2.03 M lines to 9.5 K lines. Using the program reduction techniques,  the fault signatures used to group crashes range between 52 LOC to 340 LOC\@. Such fault signatures provided fault localization information and potentially help developers focus on a small set of statements for bug understanding and diagnosis.

\subsection{Threats To Validity}\label{subsec:Threats}
\noindent{\bf Internal Threat to Validity}: One of the important challenges of evaluating deduplication of fuzzing results is that we need to have ground truth for grouping. To simulate this setting, our approach is to take known bugs and configure the fuzzers to generate only crashes for the known bugs. This approach can detect whether we are able to group crashes of the same bug together. Meanwhile, to evaluate that we do not mistakenly group crashes from one bug to other groups, we mixed all the crashes from known bugs and see whether the grouping is correct. We also consider the fact that using AFL ``Crash Exploration Mode'' may generate additional crashes from unknown bugs. We used the developers' patches to fix each bug and observe if the crashes disappear. We also used a similar approach to validate if there are any misclassification in the groups generated by \helium{} and other baselines.

\noindent{\bf External Threat to Validity}: To evaluate if our techniques can be generally applicable in practice, we used 15 different bugs from 4 real-world large open-source projects. These projects range from 18 KLOC to 313 KLOC and cover a variety of software, e.g., text based web browser and audio library. We also selected 3 SOTA widely used fuzzers and their 5 total different settings as baselines to understand if our approach indeed advances the SOTA\@. Although more crashes, bugs, software, and baselines can be useful for further confirming the generality of our approaches, our current results do provide confidence that our approach is promising and can be useful.

\section{Related Work}\label{sec:Related}

The SOTA fuzzers~\cite{zalewski2017american,Rawat_Jain_Kumar_Cojocar_Giuffrida_Bos_2017,Bohme_Pham_Nguyen_Roychoudhury_2017,certBFF,swiecki2017honggfuzz,Gan_Zhang_Qin_Tu_Li_Pei_Chen_2018,Cha_Woo_Brumley_2015,slicing-2015} use either a coverage based~\cite{zalewski2017american,Gan_Zhang_Qin_Tu_Li_Pei_Chen_2018,Bohme_Pham_Nguyen_Roychoudhury_2017} or call stack based~\cite{certBFF,swiecki2017honggfuzz,Cha_Woo_Brumley_2015,Rawat_Jain_Kumar_Cojocar_Giuffrida_Bos_2017,slicing-2015} heuristics to determine uniqueness of the fuzzed crashes and report the deduplicated fuzzed crashes.  Boehme et al.~\cite{Bohme_Pham_Nguyen_Roychoudhury_2017} extended AFL to direct the fuzzing towards a specific target, while Gan et al.~\cite{Gan_Zhang_Qin_Tu_Li_Pei_Chen_2018} improved AFL to more uniquely determine the branch coverage when fuzzing. Even though both carried out additional (manual) verification when reporting ``unique'' bugs, they didn't change AFL's underlying deduplication method of using branch coverage. Similarly, most of the call stack based approaches used the same hashing method proposed by Molar et al.~\cite{Molnar_Li_Wagner_2009} with varying number of calls (frames) used for the hashing. Of particular interest are \texttt{SYMFUZZ}~\cite{Cha_Woo_Brumley_2015} that uses ``safe stack hash'', that only considered non-corrupted call stacks, and \texttt{VUzzer}~\cite{Rawat_Jain_Kumar_Cojocar_Giuffrida_Bos_2017} that uses the last 10 basic blocks along with call stack to generate hashes to prevent classifying fuzzed crashes from different bugs into the same group. These deduplication methods are tightly coupled to their respective fuzzers and hard to use independently. In contrast, our method is agnostic of the method used for generating the fuzzed crashes. Since we capture the root cause of the bug in the fault signatures, we are also less prone to the over counting and misclassification of ``unique'' bugs found in the coverage and stack hash based methods~\cite{Klees_Ruef_Cooper_Wei_Hicks_2018}.

There are also grouping methods developed independent of the fuzzers~\cite{Chen_Groce_Zhang_Wong_Fern_Eide_Regehr_2013,Holmes_Groce_2018,van_Tonder_Kotheimer_Le_Goues_2018,Pham_Khurana_Roy_Roychoudhury_2017}. The closest related work is by Chen et al.~\cite{Chen_Groce_Zhang_Wong_Fern_Eide_Regehr_2013} that calculated a ``distance'' between fuzzed crashes to capture their static and dynamic properties and used a machine leaning to rank them. Homes et al.~\cite{Holmes_Groce_2018} and van Tonder et at.~\cite{van_Tonder_Kotheimer_Le_Goues_2018} grouped fuzzed crashes based on their responses to the mutations of the program. They hypothesize that if the behavior of two fuzzed crashes change similarly (change from crashing to not crashing) due to the same mutation, then they are more likely to be the same bug.  Pham et al.~\cite{Pham_Khurana_Roy_Roychoudhury_2017} proposed using symbolic constraints on input paths to group fuzzed crashes. They have limited applicability due to the reliance on symbolic execution to generate the constraints. Cui et al.~\cite{Cui_Peinado_Cha_Fratantonio_Kemerlis_2016} and Molar et al.~\cite{Molnar_Li_Wagner_2009} proposed call stack similarity to group fuzzed crashes. They are more prone to misclassification as discussed above. We have considered these related work as baselines, however, they are either tailored for specific applications like~\cite{Chen_Groce_Zhang_Wong_Fern_Eide_Regehr_2013, Holmes_Groce_2018} or limited in the types of bugs~\cite{van_Tonder_Kotheimer_Le_Goues_2018} or benchmarks~\cite{Pham_Khurana_Roy_Roychoudhury_2017} they can handle, while others~\cite{Cui_Peinado_Cha_Fratantonio_Kemerlis_2016,Molnar_Li_Wagner_2009} are very similar to  our current baselines.

There have also been work on performing fault localization for fuzzers. Blazytko et al.~\cite{Blazytko_AURORA_2020} is a representative work in this area. Similar to our approach of generating crash corpus, they used a known crashing input from a fuzzed crash to generate similar inputs to observe the dynamic state of the program. These are used to generate predicates similar to the input path constraints generated by symbolic execution to isolate the root cause. Variations of delta debugging~\cite{Groce_Alipour_Zhang_Chen_Regehr_2014,Christi_Olson_Alipour_Groce_2018,Vince_Hodovan_Kiss_2021,Xuan_Monperrus_2014} have also been used for fault location. Both Christi et al.~\cite{Christi_Olson_Alipour_Groce_2018} and Vince et al.~\cite{Vince_Hodovan_Kiss_2021} proposed reducing the fuzzed crashes, similar to our approach in generating fault signatures, before trying to localize bugs in order to improve their accuracy. The main difference to our work is that they only identify or rank root causes for a crash, but do not generate executable fault signatures and use them to group fuzzed crashes.

\section{Conclusions and Future Work}\label{sec:Conclusion}
This paper presents a heuristics based approach for deduplicating fuzzed crashes. As opposed to the use call stacks, code coverage, and failure symptoms based approaches, our approach uses \textit{fault signatures} to group fuzzed crashes. A fault signature captures the necessary statements that allow the bugs to be reproduced. Crashes grouped based on a fault signature thus likely share the root causes and fixes. We developed an algorithm and a tool that consist of the three components, \textit{generating fault signatures}, \textit{classifying with fault signatures} and \textit{merging fault signatures}. We evaluated our approach on 3020 fuzzed crashes against the ground truth we set up from 15 real-world bugs and patches and from 4 different open-source projects. Our results show that our approach correctly grouped 99.1\% of 3020 fuzzed crashes and generated 17 groups for 15 bugs. Our approach significantly outperformed the deduplication methods offered by 3 SOTA fuzzers, namely AFL, BFF and Honggfuzz, which reported 40--1276 groups. Considering diagnosing a crash can be challenging and time-consuming, we believe our tool can significantly improve the debugging productivity for fuzzing. In the future, we will explore the further usage of fault signatures for fault localization and automated patch generation/verification. We will also experiment our approach for grouping fuzzed crashes from different program versions and from different fuzzers.

\begin{acks}
We thank the anonymous reviewers for their valuable feedback. We also thank Xiuyuan Guo for helping with running the baseline fuzzers. This research is supported by the US National Science Foundation (NSF) under Award 1816352.
\end{acks}

\bibliographystyle{ACM-Reference-Format}
\bibliography{ms.bbl}

\end{document}